# Model calculations of a two-step reaction scheme for the production of neutron-rich secondary beams

K. Helariutta[a,*], J. Benlliure[b], M. V. Ricciardi[a], K.-H. Schmidt[a]

[a] *Gesellschaft für Schwerionenforschung, Planckstr. 1, 64291 Darmstadt, Germany*

[b] *Universidad de Santiago de Compostela, 15706 Santiago de Compostela, Spain*

**Abstract:** A two-step reaction scheme for the production of extremely neutron-rich radioactive beams, fission followed by cold fragmentation, is considered. The cross sections of the second step, the cold fragmentation of neutron-rich fission fragments, are estimated with different computer codes. Discrepancies between an empirical systematics and nuclear-reaction codes are found.

## 1. Introduction

Important progress has been achieved in experimental studies of exotic nuclei, since secondary beams of short-lived nuclear species became available. Actually, the design of more powerful next-generation secondary-beam facilities is being intensively discussed. The main challenge is the production of neutron-rich isotopes, because the neutron-drip line has only been reached for the lightest elements. The traditional way for producing neutron-rich nuclei is fission of actinides. Another approach introduced recently, based

---

[*] Present address: Laboratory of Radiochemistry, P.O. Box 55, 00014 University Helsinki, Finland

on cold fragmentation [1], has successfully been used to produce a number of new neutron-rich isotopes. Cold fragmentation seems to be best suited for producing very heavy neutron-rich nuclides which cannot be obtained by fission. In the present work, we follow the idea to combine these two methods, fission and cold fragmentation, in a two-step reaction scheme. Medium-mass neutron-rich isotopes are produced with high intensities as fission fragments. They are used as projectiles in a second step to produce even more neutron-rich nuclei by cold fragmentation.

The present work investigates the beam intensities to be realised by such a two-step reaction scheme. We concentrate our studies on the second step of this approach, the cold fragmentation of projectiles far from stability, since there are no experimental data available, while the nuclide production by fission seems to be investigated better.

Three different computer codes, EPAX, ABRABLA and COFRA, were utilised to get predictions for nuclide yields in high-energy fragmentation reactions. EPAX is a semi-empirical parameterisation of fragment cross sections [2], whereas ABRABLA [3] and COFRA are modern versions of the abrasion-ablation model. ABRABLA is a Monte-Carlo simulation code, describing the nuclear-collision process for energies well above the Fermi energy. The cold fragmentation code COFRA, which is described in ref. [1], is a simplified, analytical version of ABRABLA, which only considers neutron evaporation from the pre-fragments formed in the abrasion stage. Thus, it works only in those cases where the probability for the evaporation of charged particles is much smaller than the neutron evaporation probability. In this report, the cross-section calculations have been performed by default using the ABRABLA code. They were extended to the low cross-section values on the very neutron-rich side utilising the analytical COFRA code.

The EPAX description has carefully been adjusted to available experimental data. It well reproduces the recent cold-fragmentation data of ref. [1]. However, it is not clear, whether the predictions for the fragmentation of nuclei far from stability are realistic, since there are no experimental data available. One might hope to get more reliable predictions for these cases from a theoretical model like the ABRABLA code which includes the variations of nuclear properties like binding energies with neutron excess and their influence on the production mechanism.

## 2. Model description of cold fragmentation

Peripheral nuclear collisions at relative velocities well above the Fermi velocity can be considered in a participant-spectator picture. Nucleons in the overlap zones of projectile and target collide with each other; they are the participants. The other nucleons of projectile and target, respectively, are not directly affected by the reaction and proceed moving almost undisturbed as spectators of the reaction [4]. The main properties of the pre-fragment, formed by the projectile spectators, are the mass, the neutron-to-proton ratio, the excitation energy, and the angular momentum. ABRABLA [3] is a modern version of the abrasion-ablation model which is based on the participant-spectator picture. It makes the following quantitative predictions for the properties of the pre-fragments: The mass is directly related to the impact parameter by geometrical relations, since the number of nucleons removed is given by the volume of the projectile being sheared off by the target nucleus [5]. For a given mass loss, the protons and neutrons are assumed to be removed randomly from the projectile. The neutron-to proton ratio of the projectile is subject to statistical fluctuations as given by the hyper-geometrical distribution [6]. The excitation energy is basically given by the energies of the holes in the single-particle level scheme of the projectile after the collision [3]. Additional energy transfer from the participant zone is considered. This contribution, which is about as large as the energy of the holes, has been deduced from experimental data on very peripheral collisions [7]. The angular momentum of the pre-fragment is calculated as the sum of the angular momenta of the nucleons removed in the collision [8]. In a later stage, the pre-fragment forms a compound nucleus which consecutively evaporates particles or fissions. This de-excitation phase is calculated with an evaporation code [9]. The Glauber picture used in the abrasion model is expected to be valid at high projectile energies (above a few hundreds of MeV per nucleon). At lower energies, the transfer of nucleons sets in, leading to deep-inelastic transfer, quasi-fusion or fusion reactions [10, 11]. The validity range of the codes will be discussed later in more detail.

Since we are interested in the production of extremely neutron-rich nuclides, we will discuss the variation of the neutron-to-proton ratio in some detail. Figures 1 and 2 illustrate the calculated distributions in neutron excess and in excitation energy of the pre-fragments formed in the fragmentation of $^{197}$Au as an example. As the spatial distribu-tions of protons and neutrons are very similar, the mean value of the *N*-over-*Z* ratio of the pre-fragments is close to that of the projectile. However, the hyper-geometrical distribution predicts an important fluctuation. The most neutron-rich pre-fragments are produced, if only protons are removed. The probability for this extreme case decreases strongly with increasing mass loss. Most of the pre-fragments are highly excited. They predominantly evaporate neutrons and thus loose part of their neutron excess. Extremely neutron-rich nuclides are produced only in a cold-fragmentation process which populates the low-energy tail of the excitation-energy distribution; e.g. the proton-removal channels only survive, if the pre-fragments are formed with excitation energies below the neutron separation energy. Figures 1 and 2 demonstrate that both the probability for the abrasion of predominantly protons and the population of the low-energy tail below a given threshold decrease strongly if the number of abraded nucleons increases. These are the basic features which govern the production cross sections of neutron-rich nuclides in cold fragmentation. The results are very sensitive to the exact asymmetric, non-Gaussian shape of the excitation-energy distribution which is calculated by convoluting the energy distribution of the single-particle levels [3].

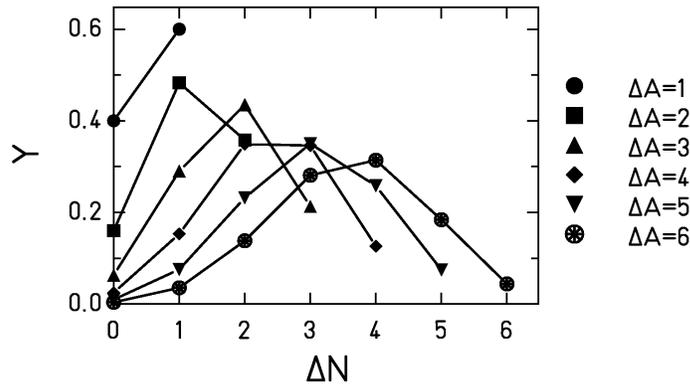

**Fig. 1. Probabilities *Y* for the removal of Δ*N* neutrons in the abrasion of 1 to 6 nucleons from $^{197}$Au, calculated using the hyper-geometrical distribution. Δ*N* = 0 means that only protons are abraded.**

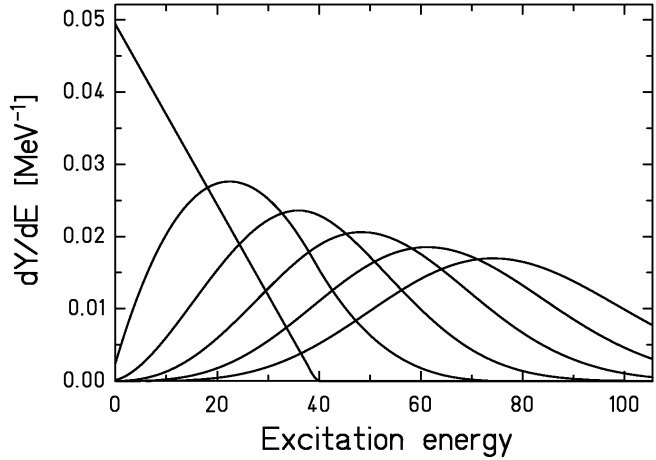

**Fig. 2. Distribution of excitation energies induced in the abrasion process by the removal of one to six nucleons (curves from the left to the right), calculated as the sum of the hole energies in the single-particle potential well [3].**

In order to favor the production of extremely neutron-rich fragments, it is certainly advantageous to start from the most neutron-rich projectile available. However, there are two effects which make it difficult to reach even more neutron-rich nuclides by cold fragmentation, if the projectile is already neutron rich. Firstly, the abrasion of neutrons is favored due to the high *N*-over-*Z* ratio of the projectile, and, secondly, the evaporation of neutrons is enhanced due to the low neutron separation energies in the neutron-rich pre-fragments. It is the main task of this work to quantitatively discuss these effects.

## 3. Results and discussion

1) <u>**Comparison of the cold fragmentation and complete abrasion-ablation model to the predictions of EPAX.**</u>

Calculations were made for three different tin isotopes ($^{112}$Sn, $^{124}$Sn and $^{132}$Sn) hitting a $^9$Be target with an energy of 1 *A* GeV. The resulting production cross sections for different fragments are shown in figures 3, 4 and 5.

For the two stable-isotope projectiles, $^{112}$Sn and $^{124}$Sn, the EPAX and ABRABLA codes seem to agree quite well. The COFRA code can not be utilised in these calcula-tions since the projectiles are too neutron deficient. With the $^{132}$Sn projectile the situation changes: Now the predictions by the two versions of the abrasion-ablation model coincide, but the results given by EPAX differ from the others. The difference of EPAX to the ABRABLA and COFRA models is increasing when moving towards the lighter elements.

The observed results could be considered from the basis of the different codes:

- EPAX is valid for stable projectiles because it is a fit to the existing data
- ABRABLA and COFRA model the physical process and thus are expected to be better suited to explore also the unknown areas.
- The cold-fragmentation code can only be utilised in the cases when the proton-evaporation probability is much less than the neutron-evaporation probability. This code seems to be well suited to describe the fragmentation of $^{132}$Sn. The full calculation with ABRABLA and the result of the cold-fragmentation code agree well in the range where both results are available. Therefore, the predictions of the cold-fragmentation code which reaches to lower cross sections can be considered as a realistic extension of the ABRABLA code for neutron-rich nuclei.

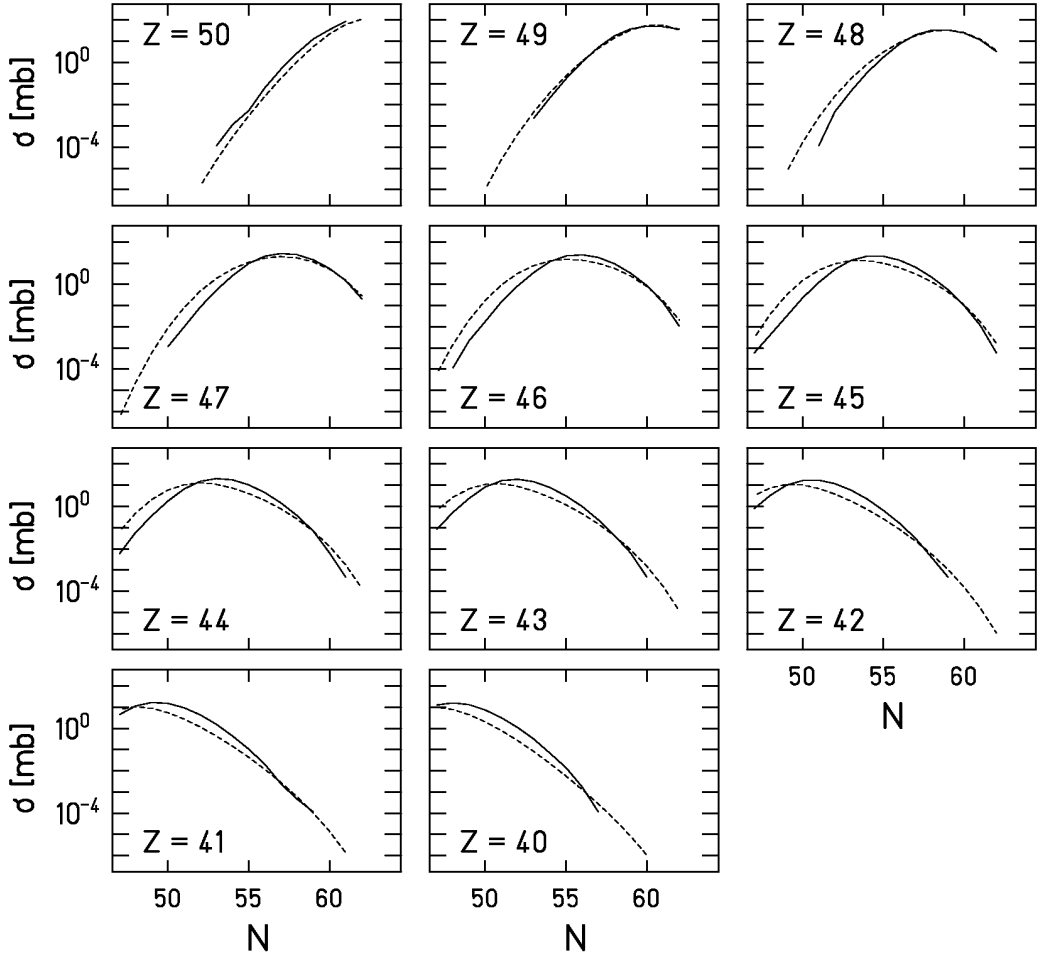

Fig. 3. Predictions for fragments from the reaction $^{112}$Sn (1 $A$ GeV) + $^{9}$Be, calculated with different codes: ABRABLA (solid line) and EPAX (dashed line).

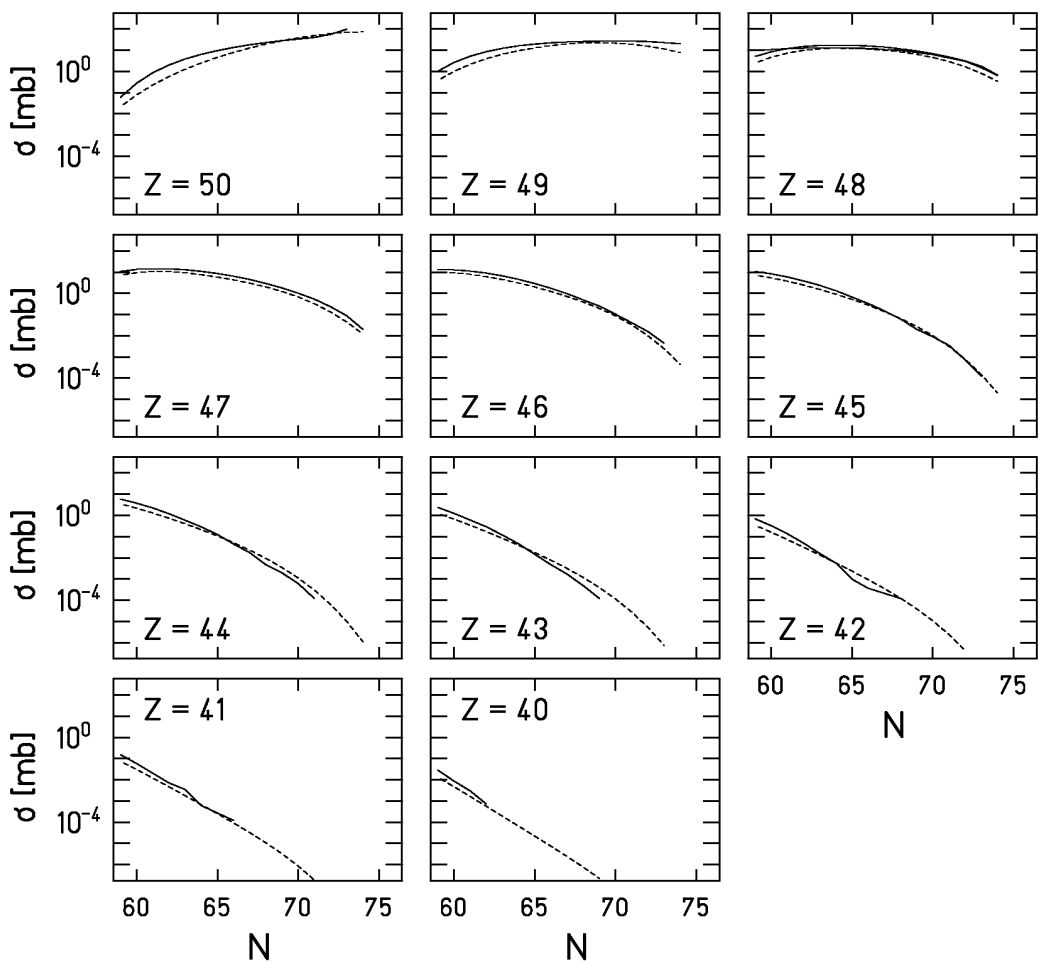

Fig. 4. Predictions for fragments from the reaction $^{124}$Sn (1 $A$ GeV) + $^9$Be, calculated with different codes: ABRABLA (solid line) and EPAX (dashed line).

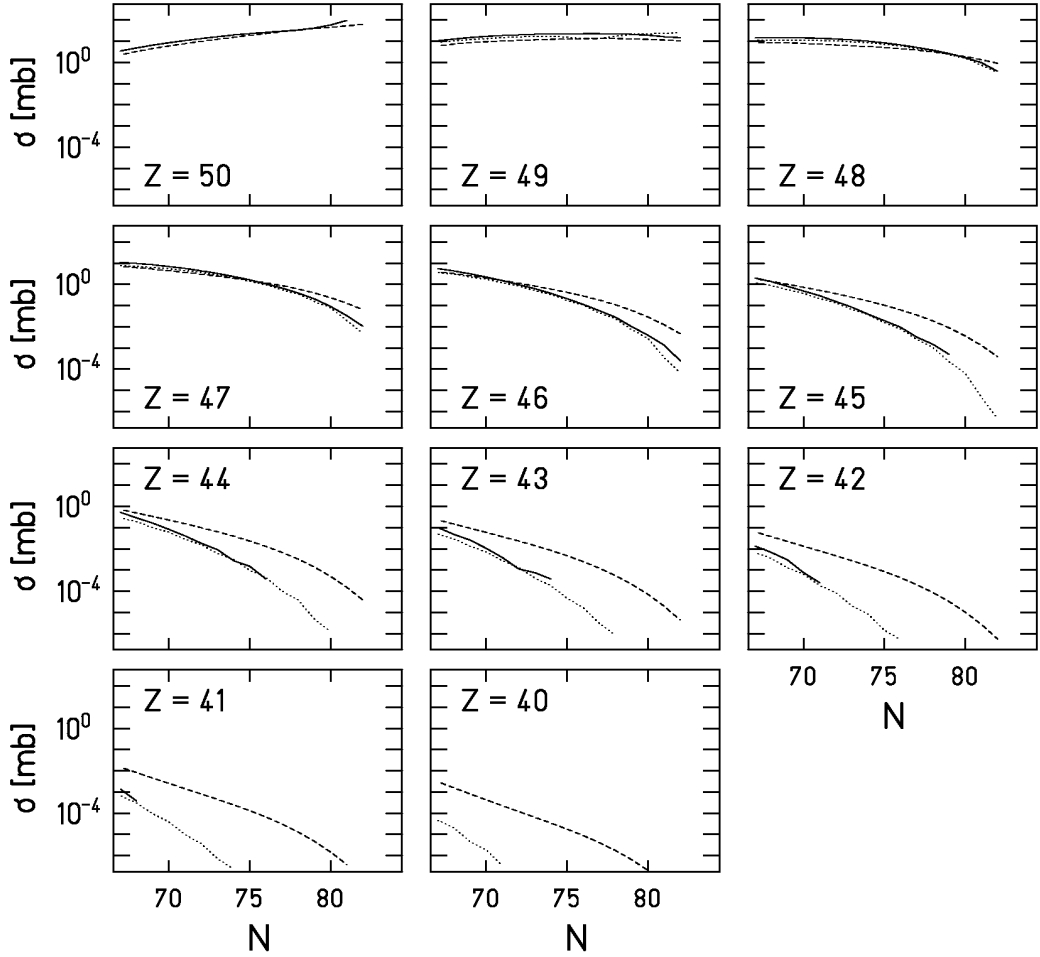

**Fig. 5.** Predictions for fragments from the reaction $^{132}$Sn (1 $A$ GeV) + $^{9}$Be, calculated with different codes: ABRABLA (solid line), COFRA (dotted line) and EPAX (dashed line).

The difference of the predictions for the general behaviour of the nuclide production in cold fragmentation of $^{132}$Sn can also be viewed on the chart of nuclides in figure 6.

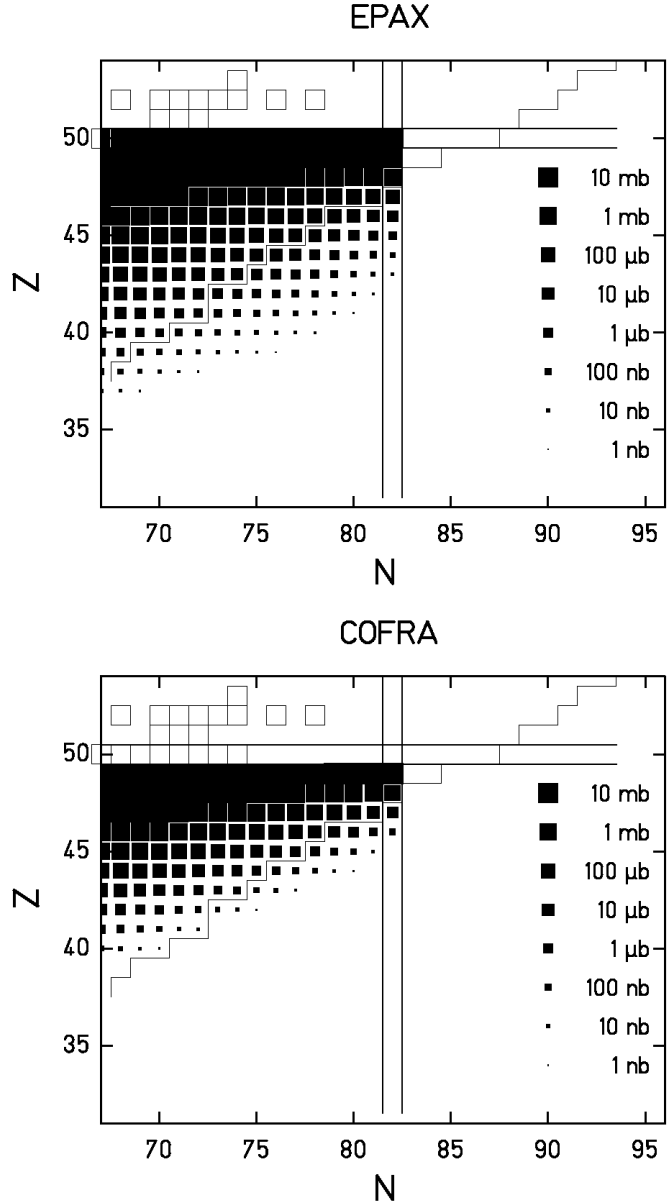

**Fig. 6.** Predicted cold-fragmentation cross sections of $^{132}$Sn in beryllium at 1 $A$ GeV from the empirical systematics EPAX and the nuclear-reaction code COFRA on a chart of the nuclides. The sizes of the clusters are a measure of the cross sections, see legend. Open squares mark the stable nuclides, while the step-like line indicates the limit of known isotopes.

We conclude that the predictions of EPAX for the fragmentation of extremely neutron-rich projectiles give much higher cross sections than the ABRABLA code, including its cold-fragmentation extension COFRA. This discrepancy sheds severe doubts on the application of EPAX to predict fragmentation cross sections using neutron-rich fission fragments as projectiles. The same precaution should be taken to apply EPAX for estimating rates from any multi-step reaction which involves neutron-rich nuclei as intermediate products as was done e.g. in ref. [12].

## 2) <u>**Comparison of the different codes to experimental data**</u>

Due to the differences of the cross sections given by the different codes, it is interesting to compare some of the calculations to measured cross-section data with special emphasis on the variation of the neutron excess. In figure 7, the experimental data on the cross sections of the most neutron-rich nuclei, produced via proton-removal channels, from different reactions are compared with the results of EPAX, ABRABLA and COFRA.

In the area of interest, all the calculated cross sections agree quite well with the available experimental results. Some trends can be seen, anyhow. The cross sections obtained with the COFRA and ABRABLA codes seem to match the data a little bit better than the EPAX cross sections. With ABRABLA it is hard to get to the very low cross sections due to the long running times, thus the cross sections are obtained only until the 3-proton removal channel. With this limited data it seems that ABRABLA would give higher cross-section values for the 4- and 5-proton removal channels compared to the COFRA code. The cross sections calculated with the EPAX code appear to underestimate the few-proton removal channels and to overestimate the many-proton removal channels.

Tests to the cross-section data from secondary reactions, i.e. the fragmentation of primary fragments on a secondary target, were performed. In this way it was possible to probe the computer codes also with unstable projectile isotopes. Figure 8 shows the comparison of the measured [13] and calculated cross sections for proton and proton-neutron removal channels for zirconium and yttrium projectiles, respectively, interacting with a beryllium target at an energy of about 1 $A$ GeV.

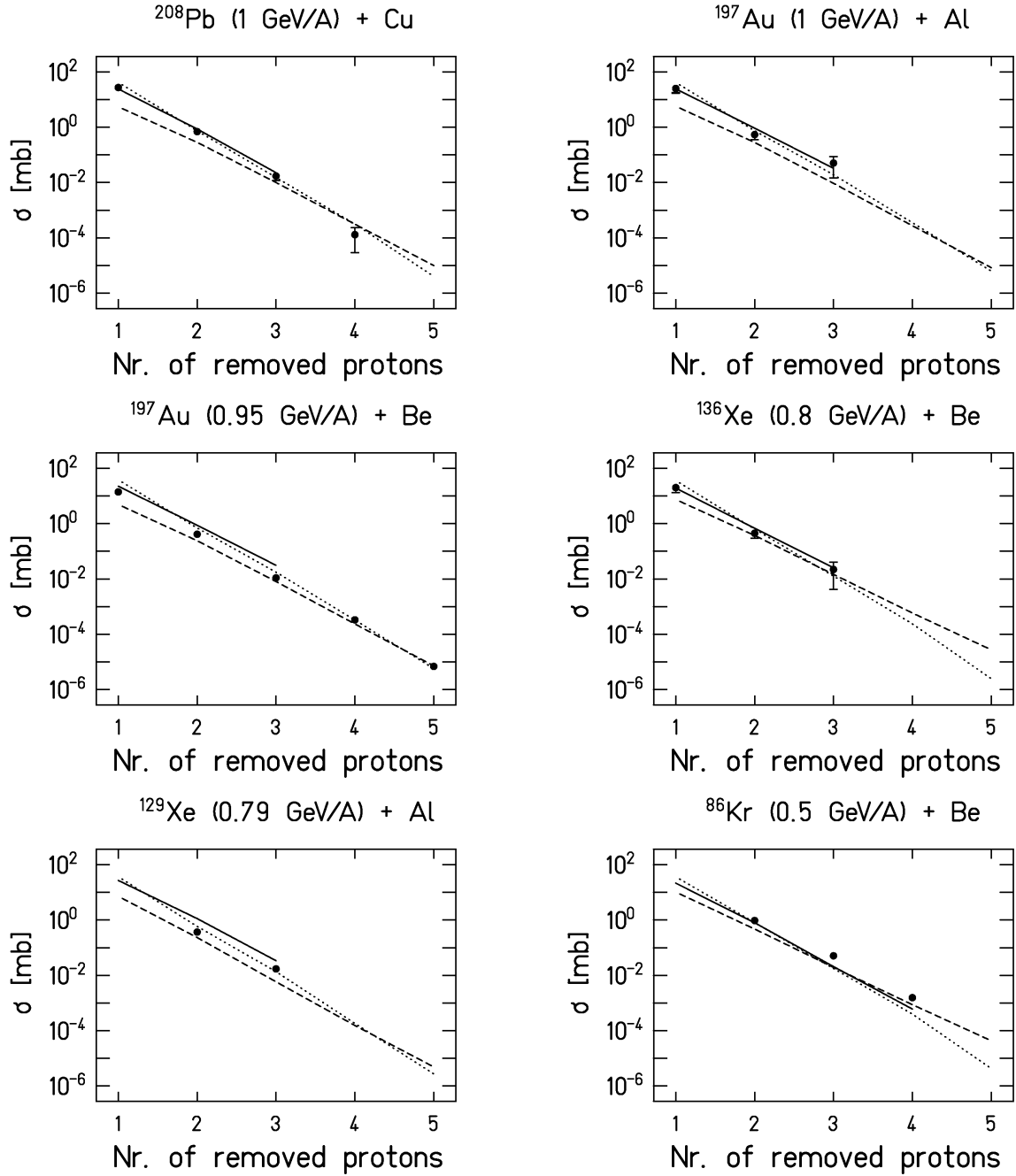

Fig. 7. The measured cross sections (closed circles) for proton-removal channels, together with the respective cross sections from calculations with EPAX (dashed line), ABRABLA (full line) and COFRA (dotted line) from the reactions $^{208}$Pb (1 $A$ GeV) + Cu [14], $^{197}$Au (1 $A$ GeV) + Al [15], $^{197}$Au (0.95 $A$ GeV) + Be[1], $^{136}$Xe (0.8 $A$ GeV) + Be [15], $^{129}$Xe (0.79 $A$ GeV) + Al [16], and $^{86}$Kr (0.5 $A$ MeV) + Be [17].

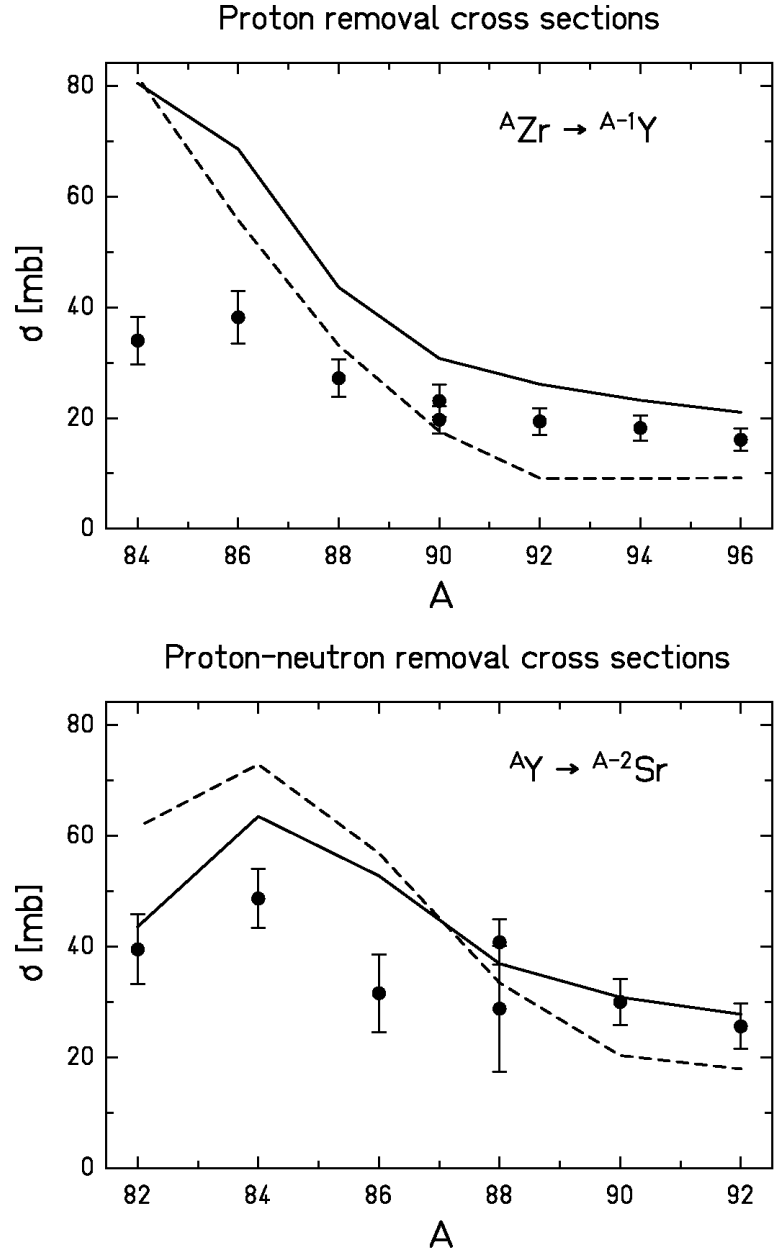

**Fig. 8.** The measured cross sections (closed circles) [13] for proton-removal channels $^{A}Zr \rightarrow {}^{A-1}Y$ and proton-neutron removal channels $^{A}Y \rightarrow {}^{A-2}Sr$, together with the respective cross sections from calculations with EPAX (dashed line) and ABRABLA (solid line) from the reactions $^{A}Zr$ (~1$A$ GeV) + Be and $^{A}Y$ (~1$A$ GeV) + Be.

Both EPAX and ABRABLA give similar kind of variation for the cross sections as a function of A, in particular a strong increase of the proton-removal channel for the most neutron-deficient zirconium projectiles. However, the experimental data do not support this kind of behaviour. We do not have any explanation for this discrepancy. The data for the most neutron-rich isotopes in which we are particularly interested seem to be slightly better reproduced by the ABRABLA code.

To test the ability to produce the gross properties of the element distributions from the fragmentation reaction of projectiles with different neutron excess, the computer codes were compared to the data from the fragmentation of two isotopes of Mn [18]. In the experiment, the $^{50,56}$Mn isotopes were interacting with a $(CH_2)_n$ target, whereas in the calculations a Be target was utilised. The use of this average target material is expected to give similar results. The calculated and experimental cross sections, summed over the various isotopes of each element, are shown in figure 9. Because the experimental data were represented in relative yields, they had to be scaled to be comparable with the calculated cross sections.

From figure 9 one can see that the experimental data show an even-odd effect which is not reproduced by any of the computer codes. While EPAX does not show any even-odd structure, the tiny enhancement in the production of even elements predicted by ABRABLA is much too small. Both EPAX and ABRABLA produce quite nicely the general trends of the element cross sections. For the $N = Z$ nucleus $^{50}$Mn, the cross-section distribution as a function of proton loss is quite flat, whereas for the more neutron-rich $^{56}$Mn, lying on the neutron-rich side of the valley of stability, the cross sections for the elements close to the projectile are significantly higher than for the lighter elements. The difference in the distributions is a consequence of the higher amount of neutrons in $^{56}$Mn than in $^{50}$Mn and thus the larger variety of isotopes produced via neutron evaporation in the vicinity of the $^{56}$Mn projectile.

We conclude that the influence of neutron excess of the projectile on the behaviour of the fragmentation cross sections is not explored sufficiently well by the available data in order to allow for an experimental verification of the differences found in the predictions of the different codes.

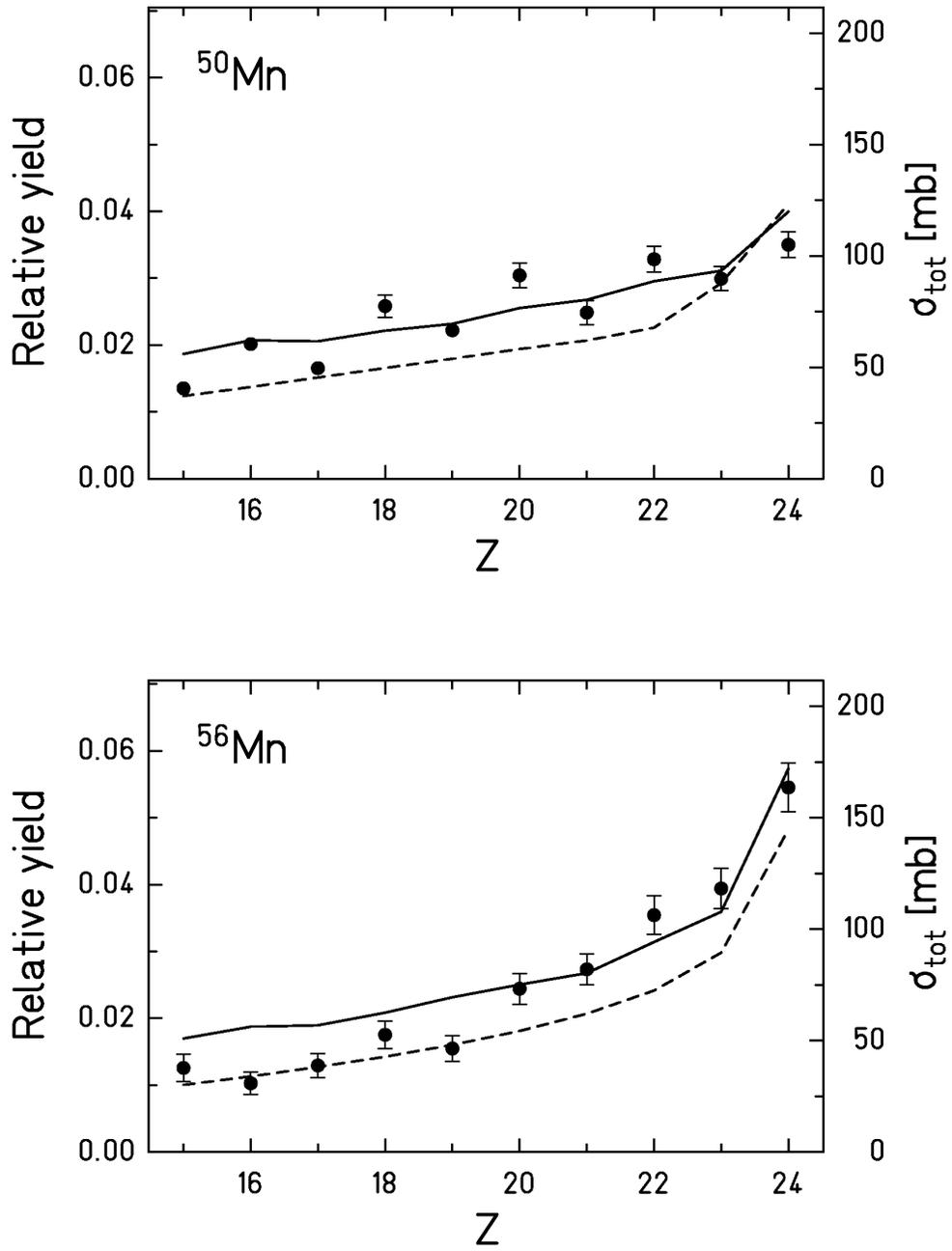

**Fig. 9. The cross sections summed over different isotopes of the elements produced in the fragmentation of $^{50}$Mn and $^{56}$Mn. The experimental data of relative yields from ref. [18] (full circles) are compared with the cross sections obtained by the ABRABLA (solid line) and EPAX (dashed line) codes. The scaling of the two different y-axes is chosen in such a way that the results can be qualitatively compared in the same figure.**

## 3) Validity of the codes at lower energies

The beam energy needed for the second step of our two-step reaction scheme is another important parameter to be investigated. First, the applicability of codes like EPAX and ABRABLA for calculating the isotopic distribution of the fragmentation products might be doubted, if the energy is too low. Secondly, the beam energy is decisive for the maximum target thickness to be used due to the electronic energy loss. In this section, we address the first problem, while the second one is discussed in the following section.

No systematic data on nuclide distributions from low-energy ($E_{projectile} \approx 75 - 200\ A$ MeV) fragmentation are available. Here the cross sections predicted by the ABRABLA and EPAX codes are compared with the data from the reactions $^{78}$Kr (75 $A$ MeV) + $^{58}$Ni [19], $^{86}$Kr (70 $A$ MeV) + Al [20] and $^{12}$C (135 $A$ MeV) + Cu [21]. The comparisons are presented in figures 10 to 12.

From the figures 10 and 11 one can observe that the ABRABLA code tends to underestimate the cross sections on the neutron-deficient side of the isotopic distribution and to overestimate them on the neutron-rich side in most cases. These tendencies can also be seen in figure 12 where all the cross sections for the isotopes on the neutron-deficient side or in the valley of stability ($^{44}$Sc, $^{52,54}$Mn, $^{52}$Fe, $^{58,60}$Co, $^{64}$Cu) are underestimated by factors of 2 to 6, whereas the cross sections on the neutron-rich side ($^{42}$K, $^{46,47,48}$Sc) are almost correctly reproduced or somewhat overestimated by the ABRABLA code. The isotopic distributions predicted by the EPAX code are rather similar, except that they are slightly shifted to the neutron-deficient side. The difference between the codes and the experiment may have two reasons: In the reactions with lower energies, mass transfer can lead to a different N/Z distribution of the prefragments, and a higher excitation energy transferred to the system will enhance neutron evaporation. However, the trends are not univocal when surveying other available data [22, 23]. Recent results on the fragmentation of $^{86}$Kr even revealed en enhanced production of neutron-rich isotopes [24]. In general, the data basis on isotopic production cross sections of reaction residues in the energy range from 75 to 200 $A$ MeV appears to be insufficient. Therefore, it seems that our model calculations can only be used with some precaution at energies below 200 $A$ MeV.

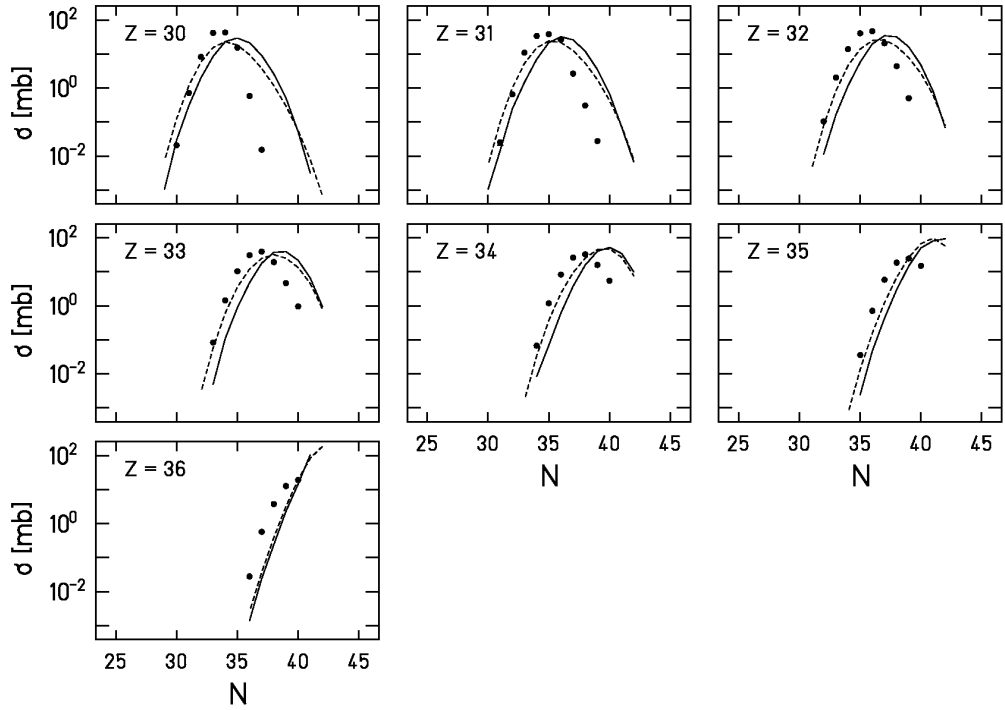

Fig. 10. The cross sections predicted by the ABRABLA (solid line) and EPAX (dashed line) codes, together with the experimental data (filled circles) obtained from the reaction $^{78}$Kr (75 $A$ MeV) + $^{58}$Ni [19].

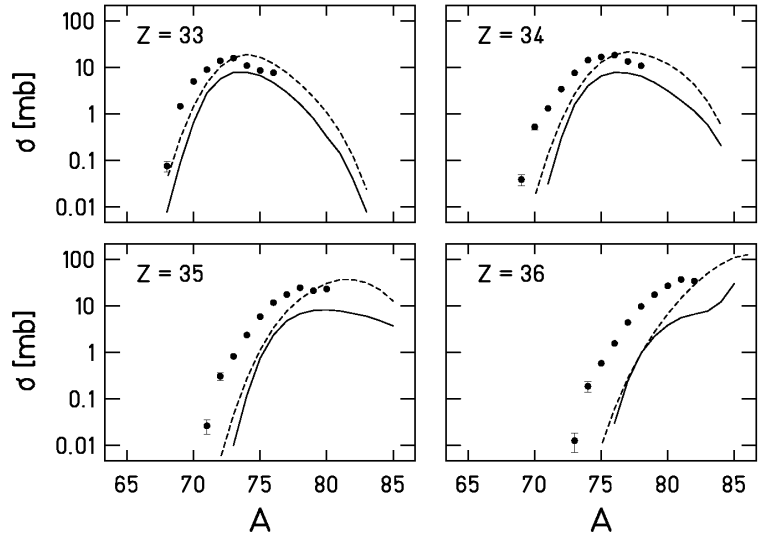

Fig. 11. The cross sections predicted by the ABRABLA (solid line) and EPAX (dashed line) code, together with the experimental data (filled circles) obtained from the reaction $^{86}$Kr (70 $A$ MeV) + Al [20].

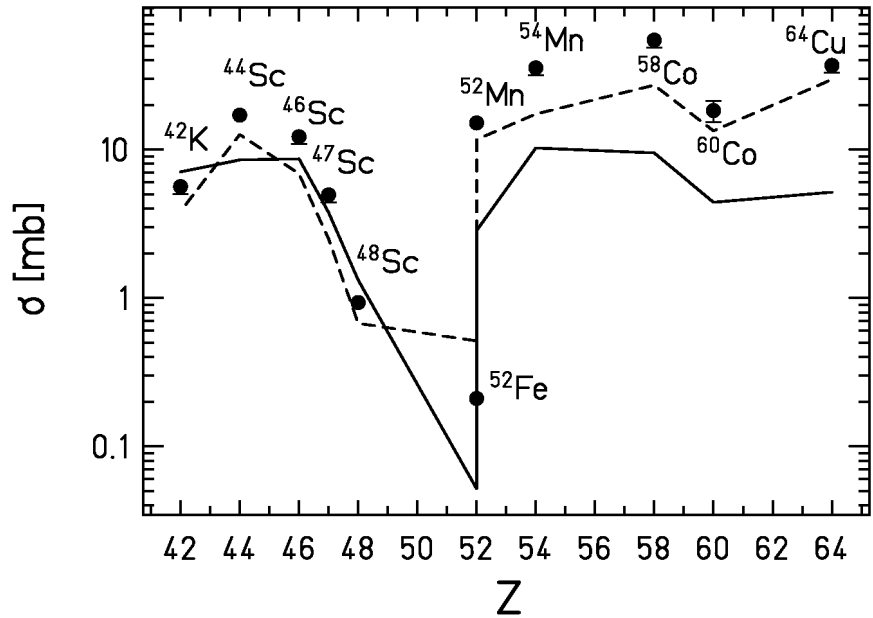

**Fig. 12. The cross sections predicted by ABRABLA (solid line) and EPAX (dashed line) for the reaction Cu + $^{12}$C, together with the experimental data (filled circles) obtained from the reaction $^{12}$C (135 *A* MeV) + Cu [21].**

### 4) <u>Estimations of fragment rates from neutron-rich projectiles</u>

Some estimations for the rates of fragments produced by neutron-rich stable and radioactive projectiles were accomplished. The production of two nuclides was studied, $^{124}$Pd and $^{78}$Ni, which are the most neutron-rich isotopes of these elements observed up to now. Palladium, like several other elements in this region, is a refractory element, which is difficult to extract from the production target by the ISOL method [25]. In addition, it results from a symmetric charge split of uranium, which is only weakly produced in low-energy fission of $^{238}$U. Low-energy fission, in the sense that shell effects have a strong influence on the fission-fragment distribution, is the relevant production mechanism in electron-induced fission [26], in the deuteron-neutron-converter concept [27] and in fission induced by spallation neutrons generated by high-energy protons. In the latter two scenarios, heating of the production target by the passage of charged projectiles is avoided, allowing for very high fission rates. Nickel is also only weakly produced in low-energy fission of $^{238}$U, since it requires a very asymmetric charge split. For both nuclides, the cross sections in direct production by fission using relativistic $^{238}$U projectiles of 750 *A* MeV in a beryllium target are experimentally known [28, 29]. According to our model calculations [30], the cross sections of neutron-rich isotopes for 1 GeV proton-induced fission of $^{238}$U are very close. This is also in agreement with measured data given in table A1 the appendix. Therefore, these data may serve as a good reference for the EURISOL project [31], which actually discusses a 1 GeV proton accelerator as driver, when direct production is considered.

In table 1, cross sections for the production of $^{124}$Pd via the fragmentation of stable ($^{136}$Xe) and radioactive ($^{132}$Sn) projectiles in beryllium have been calculated using ABRABLA and EPAX. While $^{136}$Xe is the lightest primordial nuclide with 82 neutrons, $^{132}$Sn provides the same number of neutrons with 4 protons less. $^{132}$Sn is situated on the light wing of the heavy fission-fragment component of low-energy fission of $^{238}$U. Due to the doubly-magic shell closure, it profits from a strong charge polarisation. It can be produced with high rates by low-energy fission in the scenarios discussed above. Therefore, this nucleus might be an optimum choice as a secondary projectile to produce neutron-rich nuclei in the difficult range from Z ≈ 43 to Z ≈ 49 by two-step reactions. The calculated cross sections are also compared to the measured values from the fragmentation-fission reaction of $^{238}$U at 750 *A* MeV [28]. Firstly, it can be observed in table 1 that the values from the EPAX calculations are about ten times higher that those obtained with ABRABLA. Secondly, the direct comparison of cross sections seems to be in favour of the two-step process rather than the direct production via the fragmentation of $^{136}$Xe or the fragmentation-fission of $^{238}$U. Please note that the results of these codes are essentially independent of beam energy. However, their validity is based on the applicability of the participant-spectator concept, as discussed in the preceding chapter.

**Table 1.** Comparison of production cross sections for $^{124}$Pd from different reactions, obtained with ABRABLA and EPAX by fragmentation-evaporation of $^{132}$Sn and $^{136}$Xe and from experiment by fragmentation-fission reaction of $^{238}$U [28].

| Projectile | Target | ABRABLA | EPAX | EXPERIMENT |
|---|---|---|---|---|
| $^{132}$Sn | $^9$Be | 13 000 nb | 109 000 nb | |
| $^{136}$Xe | $^9$Be | 0.39 nb | 13 nb | |
| $^{238}$U | $^9$Be | | | 32 nb |

A more honest judgement, however, needs to compare *rates* to be obtained in a given scenario. Apart from the validity range of the codes, the beam energy has an important influence on the target thickness to be used. Estimations were done for three different projectile energies: 100, 200 and 400 *A* MeV. The target thickness, corresponding to 20% and 50% of the range of $^{132}$Sn in beryllium, are shown in table 2 as an example. The energy losses and ranges in the target were calculated with the program AMADEUS [32].

**Table 2.** The thicknesses of Be targets. $E_{in}$ is the energy of the $^{132}$Sn secondary projectiles.

| $E_{in}$ [A MeV] | 20% of range of $^{132}$Sn in Be [mg/cm$^2$] | 50% of range of $^{132}$Sn in Be [mg/cm$^2$] |
|---|---|---|
| 100 | 114 | 286 |
| 200 | 365 | 912 |
| 400 | 1083 | 2709 |

The production rates *r* of the fragments were calculated using the formula [32]

$$r = t_{eff} \sigma_f n_p N_A / A_t, \qquad (1)$$

where $\sigma_f$ is the production cross section of a fragment, $n_p$ is the number of incoming projectiles, $N_A$ is the Avogadro constant, $A_t$ the mass number of target atoms and $t_{eff}$, the effective target thickness

$$t_{eff} = \frac{e^{-\mu_p t} - e^{-\mu_f t}}{\mu_f - \mu_p}, \tag{2}$$

where t is the target thickness and $\mu_p$, $\mu_f$ are the absorption coefficients for the projectile and fragment, respectively. Values of $\mu_{p,f}$ can be approximated using the relation

$$\frac{1}{\mu_i} \approx 44 \frac{A_t}{(A_t^{1/3} + A_i^{1/3})^2} \quad g/cm^2 \tag{3}$$

where $A_t$ and $A_i$ ($i = p, f$) are the mass numbers of target, projectile and fragment, respectively.

The results for the production of $^{124}$Pd by $^{132}$Sn secondary projectiles are listed in table 3. We rather arbitrarily assumed that the incoming beam intensity of post-accelerated $^{132}$Sn is $10^{11}$ particles/s. The results may easily be scaled to be adapted to other conditions. The cross sections have been calculated with the ABRABLA code. The influence of the beam energy on the production rate is rather strong due to the increasing target thickness. Going up in energy from 100 to 400 $A$ MeV gives an enhancement factor of 7 to 8. The energy spread is caused by the different energy losses of secondary projectiles and final products in the sections of the target before and after the reaction.

**Table 3. The cross section ($\sigma$) predicted by ABRABLA (see fig. 5), rates, average energies ($E$) and energy spreads ($\Delta E/E$) after a Be target for $^{124}$Pd fragments using $10^{11}$/s $^{132}$Sn projectiles. Absorption of projectiles and fragments by nuclear reactions in the target are considered.**

| $^{124}$Pd | | Target: 20% of range | | | Target: 50% of range | | |
|---|---|---|---|---|---|---|---|
| $E_{in}$ [A MeV] | $\sigma$ [µb] | Rate [1/s] | E[A MeV] | $\Delta E/E$ [%] | Rate [1/s] | E[A MeV] | $\Delta E/E$ [%] |
| 100 | 24 | 1.8·10$^4$ | 86.8 | 0.63 | 4.4·10$^4$ | 66.9 | 2.3 |
| 200 | 24 | 5.5·10$^4$ | 173.8 | 0.72 | 1.3·10$^5$ | 131.8 | 2.7 |
| 400 | 24 | 1.4·10$^5$ | 346.1 | 0.80 | 2.8·10$^5$ | 260.0 | 3.0 |

To evaluate the rates obtained in table 3 we need to compare the primary production of $^{124}$Pd in the same ISOL facility which we assumed to deliver $10^{11}$ post-accelerated secondary projectiles of $^{132}$Sn per second. A rough estimate, done by scaling on the basis of measured cross sections of $^{132}$Sn and $^{124}$Pd in the fission of $^{238}$U induced in a hydrogen target at 1 $A$ GeV and in a beryllium target at 750 $A$ MeV, respectively, gives an intensity of 9×10$^7$ nuclei per second for $^{124}$Pd in direct production (see table A1). This value is definitely higher than the yields obtained in the two-step scenario listed in table 3. However, this estimate does not consider that the extraction and ionisation efficiency for $^{124}$Pd is much smaller than the one for $^{124}$Sn. This aspect will be discussed later.

Direct production of $^{124}$Pd by fragmentation of $^{136}$Xe with a cross section of 0.39 nb (figure 5) would be competitive with the considered two-step scenario using $^{132}$Sn, provided a primary-beam intensity of about 7×10$^{15}$ $^{136}$Xe projectiles per second would be available. This comparison is approximately valid, if the beam energies of $^{132}$Sn and $^{136}$Xe are the same in the two scenarios. To provide this beam with an energy of 100 A MeV, corresponding to a beam power of 15 MW, would require a very powerful accelerator; e.g. the design value of the beam power of the accelerator for the planned RIA (Rare Isotope Accelerator) project is 400 kW [33]. Also in view of the target heating, the direct production of $^{124}$Pd by $^{136}$Xe does not seem to be a realistic scenario. However, this estimate, which is based on the calculated value of 0.39 nb mentioned above, should be considered with caution, since cold fragmentation reactions with such a large mass loss have not been observed experimentally so far.

As a second case, production rates were estimated for $^{78}$Ni and for other N = 50 isotones. In this case, a more elaborate study was performed in order to determine the production rate as a function of the nature of the secondary projectile. Table 4 shows the expected rates of $^{78}$Ni when a series N = 50 isotones is used as secondary projectiles. Extraction efficiencies are assumed to be the same in all cases. Obviously, the highest rates are obtained if the projectile is chosen as close as possible to the desired product, which is $^{78}$Ni in this case. The higher fission production cross sections of the heavier isotones do not compensate the decreasing cross section of the second reaction step. Again assuming the same extraction efficiency for nickel, the direct production of $^{78}$Ni by fission of $^{238}$U with a cross section of 0.2 nb would yield a rate of 7×10$^5$ / s (see table A1), which cannot be reached by any of the two-step options.

**Table 4:** Calculated production rates per second of $^{78}$Ni in a beryllium target by using different secondary projectiles (listed in the first column), different target ticknesses and different beam energies. See text for details.

| Targ. Thick. | 20% range | | | 50 % range | | |
|---|---|---|---|---|---|---|
| E(A MeV) | 100 | 200 | 400 | 100 | 200 | 400 |
| $^{84}$Se | 0.16 | 0.54 | 1.63 | 0.41 | 1.35 | 4.1 |
| $^{83}$As | 4.9 | 16 | 48 | 12.2 | 40 | 121 |
| $^{82}$Ge | 94 | 308 | 934 | 235 | 771 | 2334 |
| $^{81}$Ga | 519 | 1707 | 5170 | 1302 | 4266 | 12927 |
| $^{80}$Zn | 1076 | 3545 | 10740 | 2700 | 8856 | 26900 |
| $^{79}$Cu | 1124 | 3686 | 11200 | 2805 | 9215 | 28005 |

Finally, we studied the influence of the neutron number of the secondary projectile, by comparing the secondary production rates of several N = 50 isotones with $^{81}$Ga, $^{82}$Ga and $^{83}$Ga used as secondary projectiles. Figure 13 reveals that in this case the use of a very neutron-rich secondary projectile tends to reduce the production rates of the N = 50 isotones considered. A possible variation of the extraction efficiency, which is not considered here, would even enhance this trend, since the efficiency is expected to decrease with increasing neutron excess due to the shorter half lives.

For the direct production of $^{78}$Ni by fragmentation of $^{86}$Kr, COFRA predicts a cross section of 3 fb. Again, this result should be considered with caution. According to this prediction, a primary $^{86}$Kr beam intensity of more than $10^{18}$ projectiles per second would be required for direct production of $10^3$ $^{78}$Ni fragments per second, which is about competitive to the two-step scenario. Such high primary-beam intensity is completely out of reach. However, cold fragmentation remains very interesting for the production of very neutron-rich nuclides outside the fission region, where the two-step fission-fragmentation scheme cannot be applied.

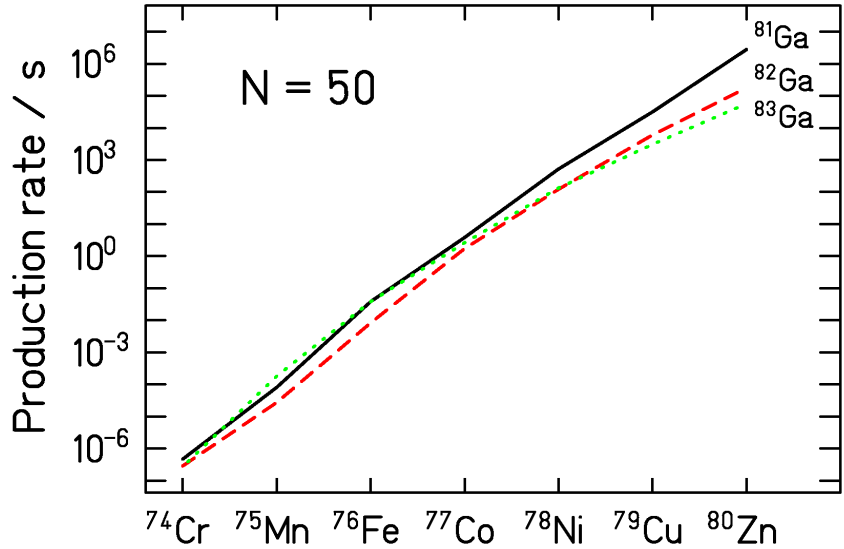

**Figure 13:** Production rates of a series of $N = 50$ isotones in a two-step reaction scheme, using $^{81}$Ga, $^{82}$Ga, and $^{83}$Ga as secondary projectiles. The beam energy is 100 $A$ MeV, and the target thickness is 20 % of the range of the respective projectile. See text for details.

We conclude that the two-step reaction scheme gives best results, when the mass loss in the second, the fragmentation step, is low. However, in the investigated cases, the one-step scenario always gives higher intensities in direct production by fission. The situation may change appreciably if we consider the available secondary-beam intensities including extraction, ionisation and re-acceleration: The two-step reaction scenario can be useful by profiting from very high secondary-beam intensities to be obtained for specific neutron-rich nuclides by the ISOL method. Extracting an abundant and long-lived neutron-rich nuclide like $^{132}$Sn from the ISOL source and fragmenting it, one can reach those isotopes that have low ISOL efficiencies due to their short half lives or due to their chemical properties [25].

The model calculations predict that direct production by fragmentation might be competitive with the two-step scenario in the far neutron-rich region, if a high-intensity pri-mary beam can be provided. This option has already been proposed previously [1] as an alternative for fission reactions in specific cases. However, these predictions may appear rather speculative, since cold-fragmentation reactions with large mass loss have not yet been investigated experimentally.

The predictions of different codes for the beam intensities to be reached in the two-step reaction scheme differ considerably. We might assume that the nuclear-reaction models ABRABLA and COFRA, which consider the variations of nuclear properties as a function of neutron excess, are better suited for extrapolating in the far neutron-rich region than the EPAX empirical systematics, which up to now exclusively relies on data from fragmentation of stable nuclei. In addition, the predictions of any of these codes at low beam energy, in particular below 200 $A$ MeV, should be validated. In any case, it would be desirable to check the diverging predictions of the different codes with a dedicated experiment. However, the requirements on such an experiment are rather high, since cross sections in the order of microbarns in the second reaction step should be reached, see figure 5. Considering previous experiments of similar kind on secondary reactions [13, 34, 35], in which cross sections above 10 mb have been investigated, this seems to be a difficult task.

## 4. Conclusions

The two-step reaction scheme, fission followed by cold fragmentation, has been investigated as a possibility to produce extremely neutron-rich isotopes. Possibilities to obtain reliable predictions for the beam intensities were considered. From a comparison with the ABRABLA code, a nuclear-reaction code, we conclude that the semi-empirical EPAX systematics gives too optimistic predictions for the residual production from fragmentation reactions of neutron-rich projectiles. Therefore, calculated cross sections for any multi-step reaction scheme, based on EPAX, may be considerably overestimated. The analytical COFRA code can be considered as a realistic extension of the ABRABLA code for neutron-rich nuclei produced with cross sections below 1 μb. Finally, the influence of the beam energy used for the second step of the reaction on the usable target thickness and on the isotopic distributions was studied. The comparison to the experi-mental data on low beam energies (70 $A$ MeV to 135 $A$ MeV) showed that the models can only be used with some precaution in this energy range.

For the cases considered in this work, the production of $^{124}$Pd and $^{78}$Ni, direct production by fission of $^{238}$U always yields considerably higher intensities than any two-step reaction scheme, if variations in the ISOL extraction efficiency are not considered. In specific cases, the two-step reaction might be a tool to profit from the high secondary-beam intensities of specific neutron-rich nuclei with favorable extraction properties from a future ISOL-type secondary-beam facility for producing beams of extremely neutron-rich isotopes of refractory elements and short-lived nuclei.

The discrepancies between the different codes indicate that the uncertainties in the theoretical predictions are important. In order to ultimately judge between the different codes, an experiment on the fragmentation cross sections of the very neutron-rich projectiles at different energies would be needed.

Cold-fragmentation reactions with large mass loss using stable beams may be considered as an interesting alternative option for the production of extremely neutron-rich isotopes. However, the predictions for this reaction type do not look encouraging compared to the two-step fission-fragmentation scheme. In addition, it lacks experimental verification.

## 5. Acknowledgement

Valuable discussions with J. Vervier which stimulated this work are gratefully acknowledged. We thank Monique Bernas for providing us with experimental cross sections of the reaction $^{238}$U + $^{1}$H before publication. This work has been performed in the frame of the EURISOL project, supported by the European Union under contract number HPRI-1999-CT-50001. Additional support by the European Community programme "Access to Research Infrastructure Action of the Improving Human Potential" under the contract HPRI-1999-CT-00001 is gratefully acknowledged.

# Appendix

## Production cross sections and beam intensities used for the calculations

The calculations performed in this work were based on a specific scenario: Radioactive nuclides are produced by bombarding a $^{238}$U target with 1 GeV protons. Radioactive nuclides are extracted from the target by the ISOL method and eventually post-accelerated to energies of a few 100 MeV. The post-accelerated secondary projectiles undergo a fragmentation reaction in a secondary beryllium target.

We arbitrarily assume that primary-beam intensity, target thickness and efficiencies lead to a post-accelerated beam of $10^{11}$ $^{132}$Sn projectiles per second. The yields of other secondary projectiles are estimated by assuming that the production cross sections scale in the same way as those measured in the interaction of $^{238}$U at 1 $A$ GeV in a hydrogen or at 750 $A$ MeV in a beryllium target and that the overall ISOL extraction efficiency is the same for all nuclides. It is further assumed that post-acceleration, including charge breeding, is performed with an efficiency of 10%.

These very crude assumptions have been made on purpose, since the efficiencies for extraction and post-acceleration constantly improve due to intense research and development. The quantitative information given in table A1 allows to adapt the numerical values given in this work to a given situation. It is also possible to adapt the values to another primary-reaction, e. g. electron-induced fission or other low-energy fission scenarios, by scaling the values given in this work with the appropriate production yields.

**Table A1: Cross sections, extracted ISOL yields and intensities of post-accelerated beams used for the model calculations in this work. The cross sections were measured in the reactions $^{238}$U + $^{1}$H at 1 A GeV [36] and $^{238}$U + $^{9}$Be at 750 A MeV [28]. The cross sections used to calculate the yields are underlined.**

| Nuclide | $^{238}$U + $^{9}$Be σ | $^{238}$U + $^{1}$H | ISOL yield | Post-accelerated |
|---|---|---|---|---|
| $^{132}$Sn | --- | <u>380 μb</u> | $10^{12}$ / s | $10^{11}$ / s |
| $^{124}$Pd | <u>32 nb</u> | | $9\times10^{7}$ / s | $9\times10^{6}$ / s |
| $^{84}$Se | 1.15 mb | <u>3.41 mb</u> | $9.6\times10^{12}$ / s | $9.6\times10^{11}$ / s |
| $^{83}$As | 503 μb | <u>1.36 mb</u> | $3.8\times10^{12}$ / s | $3.8\times10^{11}$ / s |
| $^{82}$Ge | 207 μb | <u>308 μb</u> | $1.1\times10^{12}$ / s | $1.1\times10^{10}$ / s |
| $^{81}$Ga | 22 μb | <u>34 μb</u> | $9.6\times10^{10}$ / s | $9.6\times10^{9}$ / s |
| $^{82}$Ga | <u>4.3 μb</u> | --- | $1.2\times10^{10}$ / s | $1.2\times10^{9}$ / s |
| $^{83}$Ga | <u>810 nb</u> | --- | $2.2\times10^{9}$ / s | $2.2\times10^{8}$ / s |
| $^{80}$Zn | <u>1.2 μb</u> | --- | $3.4\times10^{9}$ / s | $3.4\times10^{8}$ / s |
| $^{79}$Cu | <u>15 nb</u> | --- | $4.2\times10^{7}$ / s | $4.2\times10^{6}$ / s |
| $^{78}$Ni | <u>200 pb</u> | | $5.6\times10^{5}$ / s | $5.6\times10^{4}$ / s |